\documentclass[aps,prb,showpacs,twocolumn]{revtex4}
\usepackage{times}
\usepackage{graphicx}
\usepackage{color}

\newcommand{\ba}{\begin{eqnarray}}
\newcommand{\be}{\begin{equation}}
\newcommand{\ea}{\end{eqnarray}}
\newcommand{\ee}{\end{equation}}

\begin{document}

\title{Tunneling through magnetic molecules with arbitrary angle\\
between easy axis and magnetic field}

\author{Carsten Timm}
\email{ctimm@ku.edu}
\affiliation{Department of Physics and Astronomy, University of Kansas,
Lawrence, KS 66045, USA}

\date{February 8, 2007}

\begin{abstract}
Inelastic tunneling through magnetically anisotropic molecules is studied
theoretically in the presence of a strong magnetic field. Since the molecular
orientation is not well controlled in tunneling experiments, we consider
arbitrary angles between easy axis and field. This destroys all conservation
laws except that of charge, leading to a rich fine structure in the
differential conductance. Besides single molecules we also study monolayers of
molecules with either aligned or random easy axes. We show that detailed
information on the molecular transitions and orientations can be obtained
from the differential conductance for varying magnetic field. For random
easy axes, averaging over orientations leads to van Hove singularities in the
differential conductance. Rate equations in the sequential-tunneling
approximation are employed. An efficient approximation for their solution
for complex molecules is presented. The results
are applied to $\mathrm{Mn}_{12}$-based magnetic molecules.
\end{abstract}


\maketitle


\section{Introduction}

Electronic transport through single magnetic molecules has recently attracted a
lot of
interest,\cite{PPG02,LSB02,PaF05,ElT05,HGF06,RWH06,RWS06,TiE06,%
ElT06,JGB06,LeM06,DGR06,MiB06,GoL06,ElT07}
partly motivated by the hope for applications in molecular
electronics,\cite{Joachim,Har02,XuR06} in particular for memory devices
and quantum computation. On the other hand, the systems also show fascinating
properties of fundamental interest. Among effects observed or
predicted for transport through magnetic molecules are the Kondo
effect,\cite{PPG02,LSB02} negative differential
conductance,\cite{HGF06,RWS06} also at room temperature,\cite{ElT06}
large spin accumulation in the leads controlled by the initial spin state of
the molecule,\cite{TiE06} and large current-induced magnetization changes in
monolayers.\cite{ElT07}


Molecules with \emph{magnetic anisotropy} are ob\-served\cite{HGF06,JGB06} or
predicted\cite{RWH06,RWS06,TiE06,ElT06,ElT07} to show particularly rich
phy\-sics. Magnetic molecules based on a $\mathrm{Mn}_{12}\mathrm{O}_{12}$ core
with organic ligands (henceforce $\mathrm{Mn}_{12}$) have a large spin $S=10$
in the neutral state and strong easy-axis
anisotropy.\cite{PaP04,HGF06,JGB06,RWH06,RWS06} The anisotropy leads to an
energy barrier between states with large positive or negative spin components
along the easy (\textit{z}) axis. $\mathrm{Mn}_{12}$ has been studied
intensively with regard to quantum tunneling through this
barrier.\cite{FST96,MSS03} Recently, charge transport
through $\mathrm{Mn}_{12}$ has been studied.\cite{HGF06,JGB06,RWH06,RWS06}





Romeike \textit{et al.}\cite{RWH06,RWS06} investigate the effect of quantum
tunneling on transport through $\mathrm{Mn}_{12}$. They consider small
non-uniaxial second-order anisotropy terms $(S^x)^2-(S^y)^2 =
(S^+)^2/2+(S^-)^2/2$ and higher-order anisotropies, see also
Ref.~\onlinecite{HGF06}. The non-uniaxial terms are not present for the isolated
$\mathrm{Mn}_{12}$ molecule\cite{PaP04} but are due to the lowering of symmetry
by the leads. These non-uniaxial terms destroy the
conservation of the $z$ component of the molecular spin, $S_{\mathrm{tot}}^z$.
The molecular energy eigenstates are then not simultaneous eigenstates of
$S_{\mathrm{tot}}^z$. Since the non-uniaxial perturbation is
small, one can still work in a basis of $S_{\mathrm{tot}}^z$ eigenstates, but
incurs weak tunneling transitions between them, which are at the basis of
Refs.~\onlinecite{RWH06,RWS06}. Importantly for the present discussion, they
also study $\mathrm{Mn}_{12}$ in a magnetic field aligned with the easy axis.

The present paper starts from the realization that there is a \emph{much
stronger} symmetry-breaking effect if an external magnetic field is applied. In
present-day break-junction and electromigration experiments, the orientation of
the molecule relative to the laboratory frame is not well-controlled. However,
if the angle $\theta$ between the easy axis and the magnetic field is not small
and the Zeeman and anisotropy energies are comparable---a case that
can be realized for $\mathrm{Mn}_{12}$---there is no small parameter. The
molecular eigenstates are not approximate eigenstates of the spin component
along \emph{any} axis. This is the situation studied in the present paper.



In Sec.~\ref{sec.theory} the model Hamiltonian is introduced and the
calculation of the differential conductance is discussed. We employ a
density-matrix formalism, which allows to treat the strong electronic
interaction on the molecule \emph{exactly} and is not restricted to low bias
voltages, but treats the tunneling between molecule and leads as a weak
perturbation.\cite{Blu81,ScS94,KSS95,TuM02,BrF04,Mitra,KOO04,ElT05,KoO05,RWS06,%
TiE06,ElT06,ElT07} An approximation scheme for solving the resulting rate
equations is introduced, which works
excellently at low temperatures and leads to a large acceleration of the
numerics. Results are presented and discussed in Sec.~\ref{sec.results},
starting with Coulomb-diamond plots of differential conductance $g$ vs.\ gate
and bias voltages. Then $g$ is shown for varying magnetic field. These results
are also relevant for situations without a gate such as the STM geometry or
monolayers of molecules with parallel easy axes. Finally, the differential
conductance for monolayers of molecules with \emph{random} easy-axis
orientations is calculated. The main results are summarized in
Sec.~\ref{sec.sum}.

\section{Model and theory}
\label{sec.theory}

A magnetic molecule coupled to two conducting leads L (left) and R (right) is
described by the Hamiltonian $H=H_{\mathrm{mol}} + H_{\mathrm{L}} +
H_{\mathrm{R}} + H_{\mathrm{hyb}}$. Here, $H_{\mathrm{mol}}$ is the Hamiltonian
of the isolated molecule,
\ba
\lefteqn{ H_{\mathrm{mol}} =
  (\epsilon_0 - eV_g) \sum_\sigma c_\sigma^\dagger c_\sigma
  + U c_\uparrow^\dagger c_\uparrow c_\downarrow^\dagger c_\downarrow }
  \nonumber \\
& & {}- \Big(K_2 + \kappa \sum_\sigma c_\sigma^\dagger c_\sigma\Big) (S^z)^2
  - \mathbf{H}\cdot (\mathbf{s} + \mathbf{S})
  - J \mathbf{s}\cdot\mathbf{S} ,\quad
\label{1.Hmol2}
\ea
where $c_\sigma^\dagger$ creates an electron in the lowest unoccupied molecular
orbital (LUMO), $\mathbf{s} \equiv \sum_{\sigma\sigma'} c_\sigma^\dagger\,
(\mbox{\boldmath$\sigma$}_{\sigma\sigma'}/2)\, c_{\sigma'}$ is the spin
operator of electrons in the LUMO, and $\mathbf{S}$ is the operator of the
local spin. We also define the total spin $\mathbf{S}_{\mathrm{tot}}\equiv
\mathbf{s} + \mathbf{S}$. $\epsilon_0$ is the onsite energy of
electrons in the LUMO, which can be shifted by applying a gate voltage $V_g$,
$U$ is the repulsion between two electrons in the LUMO, and $J$
is the exchange interaction between the electron spin and the local spin. A
dependence of the anisotropy energy $K_2+\kappa n$ on the electron
number $n$ is taken into account.\cite{PaP04}



The g-factors for the local and electron spins are assumed to be equal and are
absorbed into the magnetic field $\mathbf{H}$, together with the Bohr magneton
$\mu_B$. We choose the easy axis of the molecule as the
\textit{z} axis in spin space.
Due to the rotational symmetry of $H_{\mathrm{mol}}$, the magnetic field can be
assumed to lie in the \textit{xz} plane. We neglect the small non-uniaxial
anisotropy due to the symmetry breaking by the leads and small higher-order
anisotropies in order to concentrate on the large effect due to the interplay
of anisotropy and Zeeman terms.


The Hamiltonian explicitly takes the electron in the LUMO into account, which
is non-degenerate for $\mathrm{Mn}_{12}$.\cite{PaP04} Since the next higher
energy orbital (LUMO$+1$) lies about $8\,\mathrm{meV}$ above the
LUMO,\cite{PaP04} this and higher orbitals are expected to affect the results
only weakly at the low bias voltages discussed here. Excitations to one of the
higher-energy orbitals are presumably responsible for the $14\,\mathrm{mV}$
peak observed in Ref.~\onlinecite{HGF06}.


We assume that the molecule is symmetrically coupled to the leads
$\alpha=\mathrm{L},\mathrm{R}$ by hybridization terms $H_{\mathrm{hyb}} =
\sum_{\alpha\mathbf{k}\sigma} (t\, a^\dagger_{\alpha\mathbf{k}\sigma} c_\sigma
+ \mathrm{h.c.})$, where $a^\dagger_{\alpha\mathbf{k}\sigma}$ creates an
electron in lead $\alpha$. The leads are described by non-interacting
electrons, $H_\alpha = \sum_{\mathbf{k}\sigma}
\epsilon_{\alpha\mathbf{k}\sigma}\, a^\dagger_{\alpha\mathbf{k}\sigma}
a_{\alpha\mathbf{k}\sigma}$.\cite{ElT05,TiE06,ElT06,ElT07}

There are important consequences of the presence of the transverse field $H_x$.
Since this term does not commute with the anisotropy term, $S_{\mathrm{tot}}^z$
is not conserved and the molecular eigenstates (of $H_{\mathrm{mol}}$) are not
simultaneous eigenstates of $S_{\mathrm{tot}}^z$. On the other hand, if the
magnetic field is aligned with the easy axis,\cite{TiE06,ElT06,ElT07} the
eigenvalue $m$ of $S_{\mathrm{tot}}^z$ is a good quantum number and electron
tunneling is governed by selection rules $\Delta m=\pm 1/2$ for sequential
tunneling. These selection rules do \emph{not} apply to our case. The only
selection rule still valid stems from the conservation of charge and states
that the electron number changes by $\pm 1$ for sequential tunneling. Apart
from this, all transitions are allowed. For local spin quantum number $S$,
there are $2S+1$ molecular states with $n=0$ electrons in the LUMO and
$2(2S+1)$ states with $n=1$ electrons. Thus there are $2(2S+1)^2$ transitions
between states with $n=0$ and $n=1$, leading to a much more complex
differential conductance $g$ than for the previous
case.\cite{TiE06,ElT06,ElT07}

The same model can also be used to describe a \emph{monolayer} of magnetic
molecules sandwiched between conducting electrodes, if the tunneling rate
between electrodes and molecule is sufficiently small to justify the
perturbative treatment. This can in principle be achieved by spacer groups in
the molecules or by ultrathin oxide layers. If the interactions between the
molecules can be neglected, they conduct electrons in parallel, and the
differential conductance per molecule is just the \emph{ensemble} average we
are calculating in any case. For a single molecule, we have to re-interpret the
ensemble average as a \emph{time} average over time scales long compared to the
characteristic tunneling time.

These remarks only hold for \emph{identical} molecules, which requires all
molecules to have their easy axes aligned in parallel. This is a reasonable
assumption for molecules that are deposited or assembled on the substrate with
a preferred orientation.\cite{ElT07} This has been demonstrated for mixed
$\mathrm{Mn}_{12}$ complexes containing different ligands.\cite{FCH05}
For molecules of approximately spherical shape one rather expects the
orientation of the easy axis to be \emph{random}. To study this case, we
average the differential conductance over all possible orientations.
Conversely, measurements of the differential conductance can be employed to
\emph{determine} the degree of alignment.


The absence of constants of motion other than particle number requires to
diagonalize $H_{\mathrm{mol}}$ numerically. The eigenstates typically contain
contributions from all spin $S_{\mathrm{tot}}^z$ eigenstates. The hybridization
$H_{\mathrm{hyb}}$ is treated as a perturbation, following
Refs.~\onlinecite{Blu81,Mitra,KOO04,ElT05}. This leads to a master equation for
the reduced density matrix $\rho_{\mathrm{mol}}$ in the Fock space of
$H_{\mathrm{mol}}$.

The master equation still contains off-diagonal components of
$\rho_{\mathrm{mol}}$, corresponding to superpositions of molecular
eigenstates. However, in the presence of non-commuting Zeeman and anisotropy
terms in the Hamiltonian, any two states differ in the spin expectation value
$\langle\mathbf{S}_{\mathrm{tot}}\rangle$, which leads to \emph{different}
long-range magnetic fields. Thus the unavoidable interaction
between the molecule and many degrees of freedom in the environment (e.g.,
electron spins) should impart su\-per\-se\-lec\-tion rules\cite{Zur82}
ensuring that the dephasing of superpositions is rapid. For states with
different \emph{charge}, the Coulomb field also leads to strong superselection
rules.\cite{Zur82}


It is thus sufficient to consider the rate equations for the probabilities $P_m
\equiv (\rho_{\mathrm{mol}})_{mm}$ of molecular states $|m\rangle$,
\be
\dot P_m = \sum_{n\neq m} ( R_{n\to m} P_n
  - R_{m\to n} P_m ) .
\label{1.rate2}
\ee
The \emph{stationary-state} probabilities $P^{(0)}_m$ are determined by
$\dot P^{(0)}_m = 0$.
The transition rates $R_{m\to n}$ are written as a sum over leads
and spin directions, $R_{m\to n}=\sum_{\sigma\alpha}
R^{\sigma\alpha}_{m\to n}$, with
\ba
R^{\sigma\alpha}_{m\to n} & = & \frac{1}{\tau_0} \big[
  f(\epsilon_n-\epsilon_m+e s_\alpha V/2)\, |C_{mn}^\sigma|^2
  \nonumber \\
& & {}+ f(\epsilon_n-\epsilon_m-e s_\alpha V/2)\, |C_{nm}^\sigma|^2 \big] ,
\label{C1.R4}
\ea
where $s_{\mathrm{L}}=1$, $s_{\mathrm{R}}=-1$,
$1/\tau_0 \equiv 2\pi\,
|t|^2\, Dv_{\mathrm{uc}}/\hbar$ is the typical transition rate in terms of
the density of states $D$ (for one spin direction)
of the leads and their
unit-cell volume $v_{\mathrm{uc}}$, $V$ is the bias voltage,
$\epsilon_m$ is the energy of state $|m\rangle$, $f(x)$ is the Fermi
function, and $C_{mn}^\sigma \equiv \langle m|c_\sigma|n\rangle$
are transition matrix elements.
The current $I_\alpha$ through lead $\alpha$ can be expressed in terms of the
rates and probabilities,
\be
I_\alpha =
  -e\,s_\alpha \sum_{mn\sigma} (n_m-n_n)\, R^{\sigma\alpha}_{n\to m}
  \, P_n ,
\label{C1.I5}
\ee
where $n_m$ is the number of electrons for state $|m\rangle$.
We are interested in the stationary state for which the current through both
leads is equal. Therefore, we drop the subscript $\alpha$ and insert the
stationary-state probabilities $P^{(0)}_n$. We will mainly analyze the
differential conductance $g \equiv dI/dV$.

In principle, the solution of the equations $\dot P^{(0)}_m = 0$ is simple: The
probabilities $P^{(0)}_m$ form an eigenvector of a matrix $A$ to zero
eigenvalue, where $A_{mn} = R_{n\to m}$ for $m\neq n$ and $A_{mm} =
-\sum_{p\neq m} R_{m\to p}$. However, numerical diagonalization often fails
because the components of $A$, the rates, vary over many orders of magnitude
due to the Fermi factors in Eq.~(\ref{C1.R4}). The ratio of very small rates
can have a large effect on the probabilities, in particular for anisotropic
magnetic molecules.\cite{TiE06,ElT06} Consequently, truncation errors in the
very small rates can lead to large errors in the probabilities.


An approach that avoids this problem would be highly welcome. One such approach
relies on the enumeration of all tree graphs on the network of molecular states
connected by allowed transitions.\cite{Hil66,Sch76,ZiS06} However, this
approach is not feasible for large molecular Fock spaces, since it requires
summation over of the order of $N^{N-2}$ tree graphs\cite{ZiS06} for each
$P^{(0)}_m$, where $N$ is the dimension of the space.

The solution at $T=0$ is simpler and less susceptible to truncation
errors because the Fermi factors are all either zero or unity. Thus there are
no exponentially small rates, unless some matrix elements $C^\sigma_{mn}$ are
exponentially small. This is not the case for generic angles
$\theta$ between magnetic field and easy axis. We here employ an approximation
scheme to find the $P^{(0)}_m$ and the current $I$. The scheme relies on
solving the rate equations and calculating the current \emph{exactly} at $T=0$
and introducing broadening of the steps in $P^{(0)}_m$ and $I$ to take finite
temperatures into account. Details are presented in App.~\ref{app.a}. The
approximation is excellent at sufficiently low temperatures. It speeds up the
calculation of $g$ for $500\times 500$ values of $V_g$ and $V$ and spin
$S=2$ by a factor of about 350.

\section{Results and discussion}
\label{sec.results}

In the following, we discuss results for molecules with local spin $S=2$ and
$S=10$. The smaller spin allows to exhibit the physics more clearly, since the
number of relevant transitions is smaller. As noted above, there are
$2(2S+1)^2$ transitions between states with zero and one electron, which gives
$50$ for $S=2$ and $882$ for $S=10$. These are the maximum possible numbers of
differential-conductance peaks. For $S=2$, we choose parameter values that
allow to discuss the effects of interest but do not correspond to a specific
molecule.



For $S=10$ we use realistic parameters for $\mathrm{Mn}_{12}$ calculated by
Park and Pederson\cite{PaP04} employing density functional theory in the
generalized-gradient approximation. We take $K_2 = 0.0465\,\mathrm{meV}$,
$\kappa = -0.00862\,\mathrm{meV}$ (taken from results for potassium doping),
and $J = 3.92\,\mathrm{meV}$ (from the energy difference of states with
$m=21/2$ and $m=19/2$, respectively, where $m$ is the quantum number of
$S^z_{\mathrm{tot}}$). $K_2$ is close to the experimental value
$K_2 = 0.056\,\mathrm{meV}$ from Ref.~\onlinecite{HGF06}.



The onside energy $\epsilon_0$ cannot be inferred from Ref.~\onlinecite{PaP04}.
The LUMO--HOMO gap is $E_g = 438\,\mathrm{meV}$.\cite{PaP04} Since
$\mathrm{Mn}_{12}$ is found to be more easily negatively doped, we can assume
$\epsilon_0$ to lie closer to the LUMO. In experiments with a gate, the onsite
energy is shifted to $\epsilon_0-eV_g$ in any case.  For single
$\mathrm{Mn}_{12}$ molecules, we will mainly consider the parameter region
close to the crossing point where states with $n=0$ and $n=1$ become
degenerate. Finally, we choose a very large value for $U$ so that double
occupation of the LUMO is forbidden.


\begin{figure}[t]
\centerline{\includegraphics[width=1.68in]{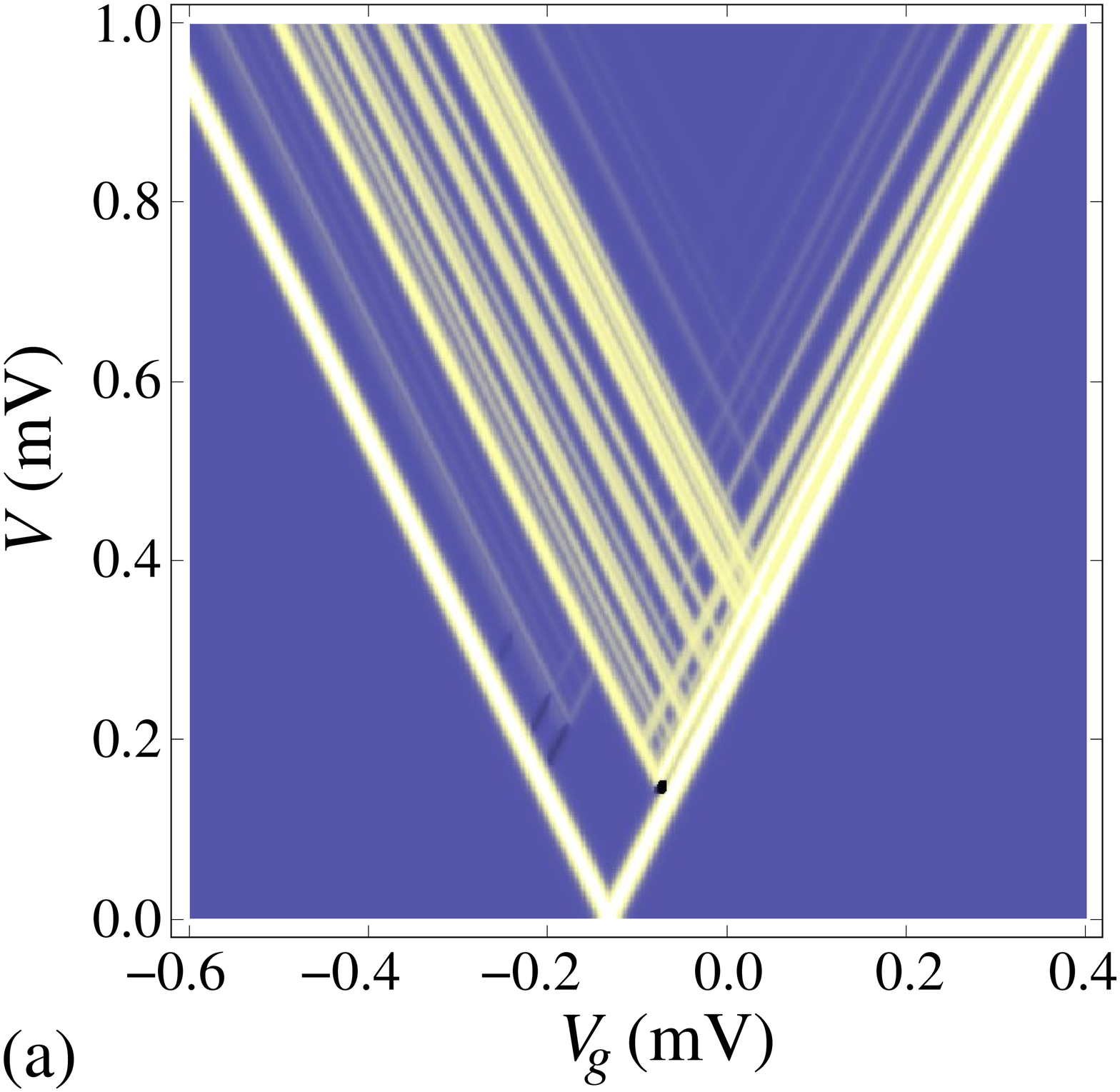}
\includegraphics[width=1.68in]{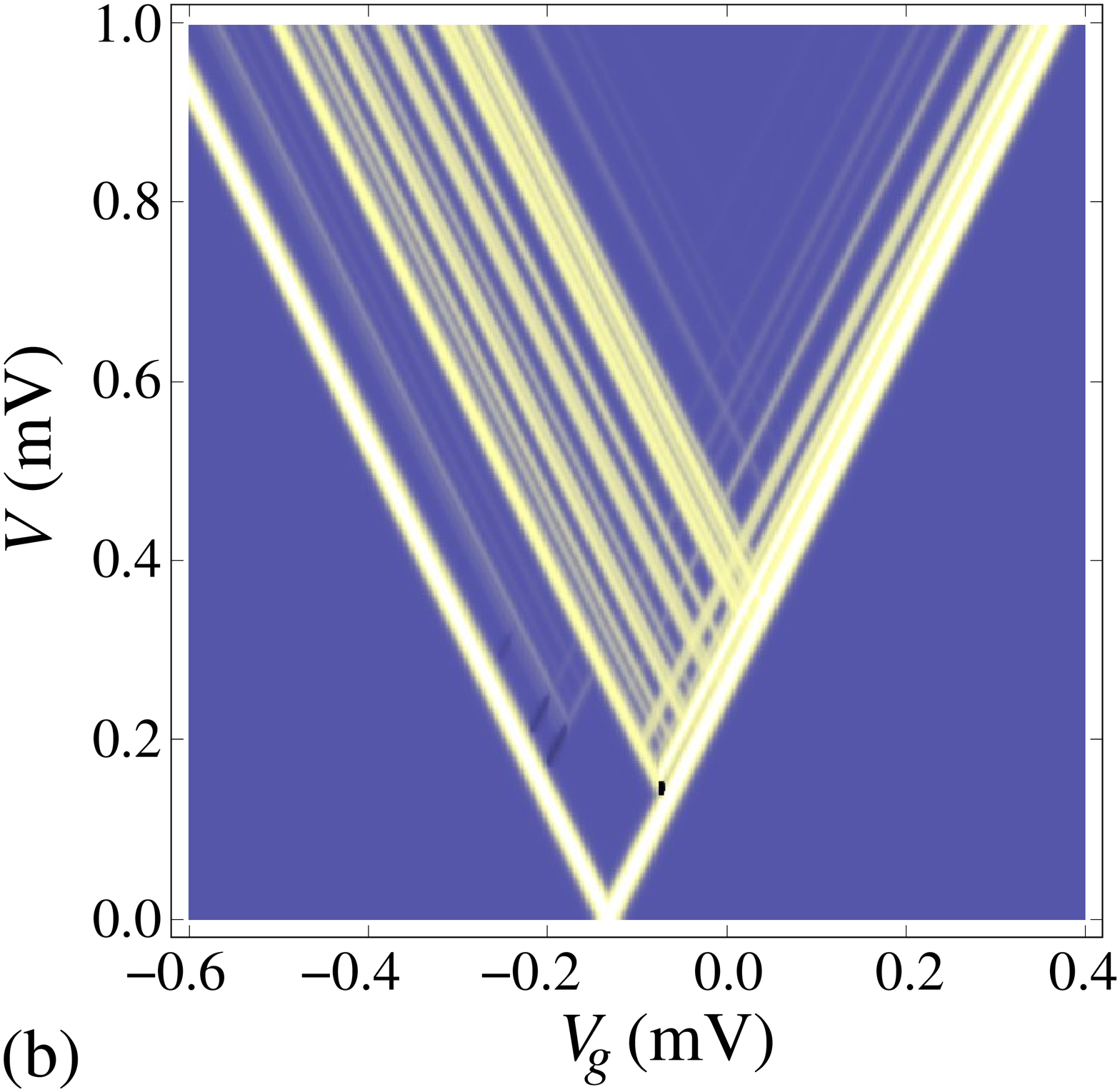}}
\caption{\label{fig.g1}(Color online) Differential conductance
$g=dI/dV$ as a function of gate voltage
$V_g$ and bias voltage $V$. Bright (dark) colors denote $g>0$ ($g<0$).
The parameters
are $S=2$, $\epsilon_0=0$, $U=10$, $H_x=H_z=0.05$, $K_2=0.04$,
$\kappa=0$, $J=0.1$, and $T=0.002$ (all energies in meV).
(a) Results obtained by exact solution of the rate equations.
(b) Results of the approximation of App.~\protect\ref{app.a}.}
\end{figure}

\subsection{Gate-voltage scans}

We start by presenting a typical differential-conductance plot as a function of
gate voltage $V_g$ and bias voltage $V$ for spin $S=2$ in Fig.~\ref{fig.g1}.
The magnetic field is applied at an angle $\theta=45^\circ$ relative to the
easy axis. The comparison of the exact solution of the stationary-state rate
equations with the approximation of App.~\ref{app.a} shows excellent agreement.


The blue (medium gray) regions to the left and right of the crossing point in
Fig.~\ref{fig.g1} correspond to Coulomb blockade (CB) with small current and
electron numbers $n=0$ and $n=1$, respectively. The plot would be mirror
symmetric for $V<0$. The CB regions are delimited by strong peaks at the CB
threshold, where the current increases rapidly to a large value of order
$e/\tau_0$. In the absence of internal degrees of freedom of the molecule this
would be the only structure in the plots. However, the interaction of the
tunneling electron with the local spin leads to \emph{inelastic} tunneling
processes, which cause the peaks at the CB threshold to split.

\begin{figure}[th]
\includegraphics[width=1.1in]{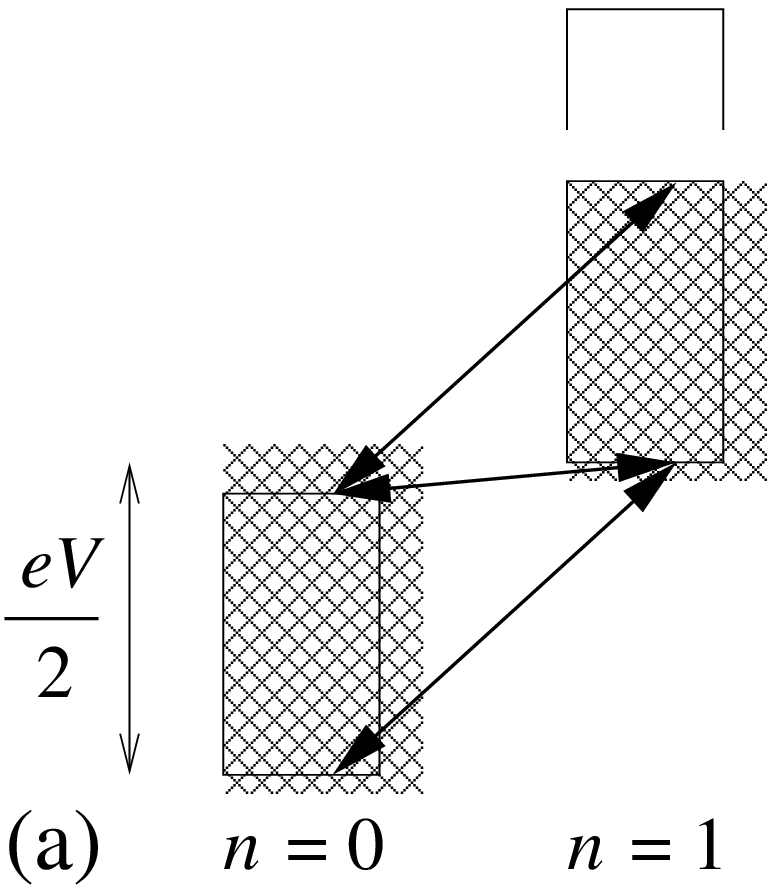}
\hspace{4em}
\includegraphics[width=1.1in]{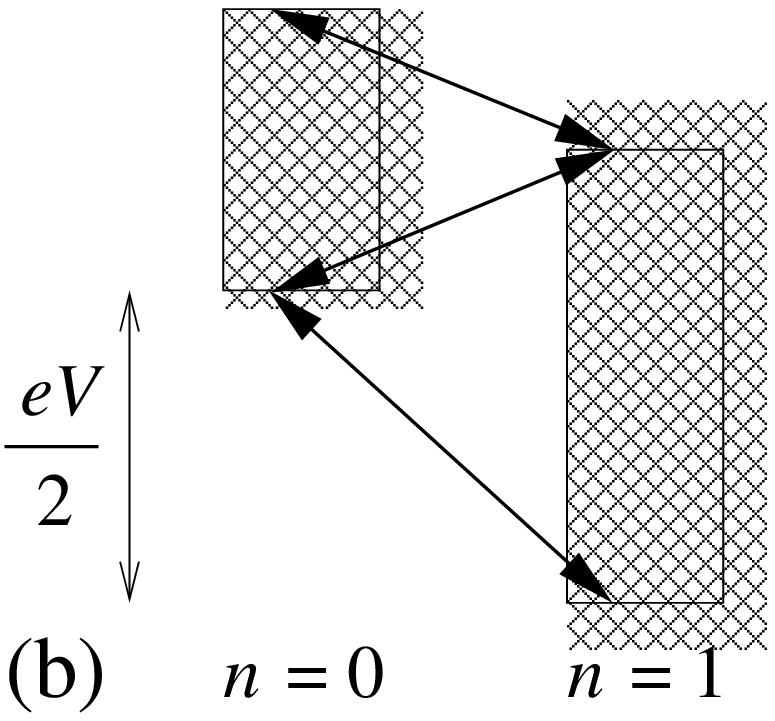}
\caption{\label{fig.multi}Sketch of molecular energy multiplets (a) to the left
of the crossing point in Fig.~\protect\ref{fig.g1} (ground state with
$n=0$ electrons) and (b) to the right of the crossing point (ground state with
$n=1$). A few representative transitions are shown as double-headed arrows.
Generically,
all transitions between states with $n=0$ and $n=1$ have nonzero matrix
elements. The
bias voltage $V$ is at the CB threshold in both cases. States that become
populated at the CB threshold at $T=0$ are cross-hatched. In
(a) some states with $n=1$ do not become active at the CB threshold (white
rectangle).}
\end{figure}

The resulting fine structure is much more complex than for magnetic molecules
without anisotropy\cite{ElT05} or in the absence of a strong magnetic
field,\cite{TiE06,ElT07} since there are many more allowed
transitions in the present model due to non-commuting Zeeman and anisotropy
terms. Each peak in $g$ corresponds to one or more allowed
transitions becoming energetically possible. The peak at the CB threshold is
\emph{much stronger} (the current step is much higher) than the peaks at higher
bias voltage, since many additional transitions become available at the CB
threshold, as illustrated by Fig.~\ref{fig.multi}.

Figure \ref{fig.g1} also shows strong \emph{asymmetry} between the fine
structure on both sides of the crossing point. This is due to the multiplet of
states with $n=1$ being \emph{broader} in energy than the one with $n=0$.
Figure \ref{fig.multi} shows that for a ground state with $n=0$, \emph{not} all
states with $n=1$ become active (obtain nonzero probability at $T=0$) at the CB
threshold, whereas for a ground state with $n=1$, all state with $n=0$ become
active. Thus for the first case, to the left of the crossing point, there are
more $g$ peaks where additional states with $n=1$ become active.

Why then are there \emph{any} additional lines to
the right of the crossing point? Here, all $n=0$ states become
active at the CB threshold. However, the probabilities and the current show
steps when additional transitions become energetically possible even if the
final state of these transitions was already active. This mechanism leads to
$g$ peaks on \emph{both} sides.





\begin{figure}[t]
\centerline{\includegraphics[width=3.20in]{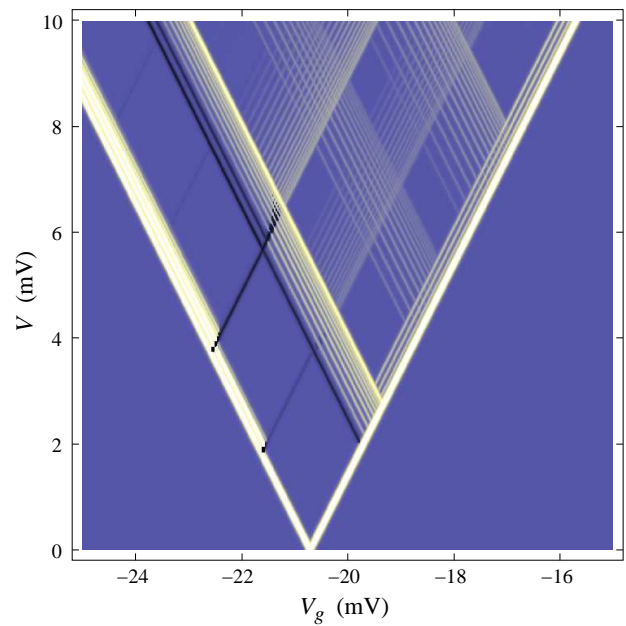}}
\caption{\label{fig.g2}(Color online) Differential conductance
$g$ as a function of gate voltage
$V_g$ and bias voltage $V$ for $\mathrm{Mn}_{12}$ in a magnetic
field perpendicular to the easy axis.
Bright (dark) colors denote $g>0$ ($g<0$). The parameters
are $S=10$, $\epsilon=0$, $U=10000$, $H_x=2.32$
(corresponding to $20\,\mathrm{T}$), $H_z=0$, $K_2=0.0465$, $\kappa=-0.00862$,
$J=3.92$, and $T=0.00863$ ($0.1\,\mathrm{K}$), where
all energies are in meV.
The approximation of App.~\protect\ref{app.a} has been used.}
\end{figure}

\begin{figure}[t]
\centerline{\includegraphics[width=1.68in]{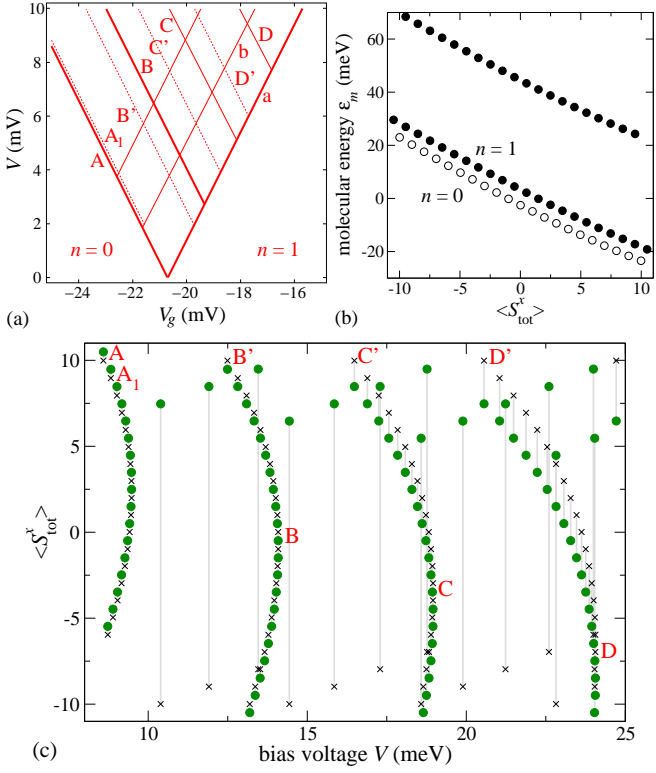}
\includegraphics[width=1.68in,clip]{timmfig4b.eps}}

\vspace*{1ex}
\includegraphics[width=3.20in,clip]{timmfig4c.eps}
%
\caption{\label{fig.g2a}(Color online)
(a) $g$ peaks corresponding to special transitions in
Fig.~\protect\ref{fig.g2}, see text.
(b) Molecular energy levels vs.\ spin expectation value $\langle
S_{\mathrm{tot}}^x\rangle$, for $\epsilon_0-eV_g=25\,\mathrm{mV}$.
Open (solid) circles
correspond to states with $n=0$ ($n=1$) electrons in the LUMO.
(c) Spin expectation values $\langle S_{\mathrm{tot}}^x\rangle$ 
of initial (crosses) and final (circles)
states vs.\ bias voltage for the lowest-lying $g$ peaks,
for $\epsilon_0-eV_g=25\,\mathrm{meV}$.
Only transitions starting from states with $n=0$ are shown.
Special transitions are labeled as in (a).}
\end{figure}

Next, we consider parameters for $\mathrm{Mn}_{12}$. Figure \ref{fig.g2}
shows $g(V_g,V)$ for a strong \emph{transverse} field ($H_z=0$) close to the
crossing point between CB with $n=0$ and $n=1$.
Note that $\epsilon_0$ has been set to zero, the zero of $V_g$ is thus
arbitrary. Figure \ref{fig.g2a}(a) shows a sketch of special features in
Fig.~\ref{fig.g2}.

The plots show two clear energy scales. For the strong magnetic field
($20\,\mathrm{T}$) in Fig.~\ref{fig.g2}, the Zeeman energy dominates over the
anisotropy energy. The molecular states are \emph{nearly} eigenstates of
$S_{\mathrm{tot}}^x$. In the following we use Fig.~\ref{fig.g2} as an example
for the analysis of differential-conductance plots for complex molecules. We
concentate on the case when the ground state has $n=0$ electrons, to the left
of the crossing point.

To facilitate the analysis, Fig.~\ref{fig.g2a}(b) shows the energy of low-lying
molecular states vs.\ the expectation value $\langle
S_{\mathrm{tot}}^x\rangle$, which is nearly a good quantum number. Figure
\ref{fig.g2a}(c) shows, for each dif\-fe\-ren\-tial-con\-duc\-tance
peak (current step) occuring at low bias voltage, the
spin expectation value $\langle S_{\mathrm{tot}}^x\rangle$ for the initial
(cross) and final (circle) state vs.\ the bias voltage. Only
transitions starting from states with $n=0$ are included, these have negative
slope in Fig.~\ref{fig.g2}. Several transitions are marked with the same
letters as in Fig.~\ref{fig.g2a}(a). These plots assume
$\epsilon_0-eV_g=25\,\mathrm{meV}$ (the left edge of Fig.~\ref{fig.g2}), but
the conclusions hold in general far left of the crossing point.


The CB threshold corresponds to transition A in Figs.~\ref{fig.g2a}(a) and (c).
Since all matrix elements between states with $n=0$ and $n=1$ are nonzero,
nearly all states belonging to  the low-energy ladder in Fig.~\ref{fig.g2a}(b)
assume nonzero probabilities even at $T=0$ as soon as transition A becomes
energetically possible.\cite{footnote.highmulti} As noted above, this makes the
threshold peak anomalously strong.


The second strong line corresponds to transition B. Without anisotropy, A and B
would be the \emph{only} visible transitions from $n=0$ to $n=1$, since
transition A (B) would correspond to a change in $\langle
S_{\mathrm{tot}}^x\rangle$ by $+1/2$ ($-1/2$) and these would be the only
allowed changes. Also, the ladders of $n=0$ and $n=1$ levels would be exactly
parallel so that all allowed transitions would have one of these two energies.
In our case, the anisotropy leads to additional allowed transitions C, D,\dots
etc.\ corresponding to changes of $\langle S_{\mathrm{tot}}^x\rangle$ by 
approximately $-3/2, -5/2, \ldots$, respectively.

The peaks show additional fine structure with a smaller energy scale coming
from the anisotropy. The CB-threshold peak is accompanied by additional peaks
at slightly higher bias. The first visible one corresponds to transition
$\mathrm{A}_1$ in Fig.~\ref{fig.g2a}(c). Peaks B, C, D are
accompanied by series of peaks at \emph{lower} bias, reaching down to B', C',
D'. These peaks correspond to arc-shaped series of transitions with
approximately the same change in $\langle S_{\mathrm{tot}}^x\rangle$. There are
a few additional transitions with large changes in $\langle
S_{\mathrm{tot}}^x\rangle$ in Fig.~\ref{fig.g2a}(c). These are not visible in
Fig.~\ref{fig.g2} due to very small transition matrix elements.

Several peaks, in particular B', show negative differential conductance (NDC).
The origin is the following:\cite{TiE06}
The current equals the charge $e$ times a typical
tunneling rate. This tunneling rate is a weighted average over the rates of
energetically possible transitions. But transition B' has a \emph{small} rate
due to small $C^\sigma_{mn}$, as numerical calculation shows. Therefore, the
typical tunneling rate and the current \emph{decrease} at B'.


\begin{figure}[t]
\centerline{\includegraphics[width=1.68in]{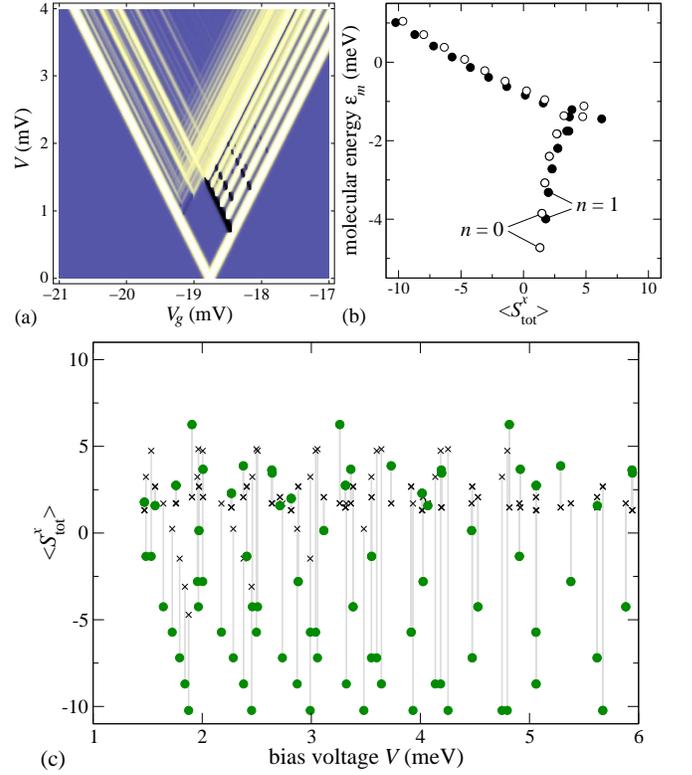}
\includegraphics[width=1.68in,clip]{timmfig5b.eps}}

\vspace*{1ex}
\includegraphics[width=3.20in,clip]{timmfig5c.eps}
%
\caption{\label{fig.g3}(Color online) (a) Differential conductance
$g$ as in
Fig.~\protect\ref{fig.g2} except for a smaller magnetic field,
$H_x=0.116\,\mathrm{meV}$ ($1\,\mathrm{T}$).
The approximation of App.~\protect\ref{app.a} has been used.
(b) Molecular energy levels vs.\ spin expectation value $\langle
S_{\mathrm{tot}}^x\rangle$, for $\epsilon_0-eV_g=19.5\,\mathrm{mV}$.
Open (solid) circles
correspond to states with $n=0$ ($n=1$).
(c) Spin expectation values $\langle S_{\mathrm{tot}}^x\rangle$ 
of initial (crosses) and final (circles) states vs.\ bias
voltage for the lowest-lying $g$ peaks, for
$\epsilon_0-eV_g=19.5\,\mathrm{meV}$.
Only transitions starting from states with $n=0$ are shown.}
\end{figure}


For comparison, Fig.~\ref{fig.g3}(a) shows the differential conductance for a
smaller transverse field leading to comparable Zeeman and anisotropy energies.
In this case, the molecular states are no longer approximate eigenstates of
$S_{\mathrm{tot}}^x$, cf.\ Fig.~\ref{fig.g3}(b). This leads to a more
complicated fine structure. It is still possible to attribute peaks to specific
molecular transitions, as comparison with the map of transitions in
Fig.~\ref{fig.g3}(c) shows, but the spectrum and the transition energies look
more random.


Note that the fine structure splitting is now broader where the ground state
has $n=1$ electrons, opposite to the $S=2$ molecule, cf.\
Fig.~\ref{fig.g1}. This stems from the smaller anisotropy $K_2+\kappa<K_2$
in the $n=1$ case, which makes the multiplet of accessible $n=1$ states
\emph{narrower} in energy than the $n=0$ multiplet, as seen in
Fig.~\ref{fig.g3}(b).\cite{footnote.Jstates}



\subsection{Magnetic-field scans}

Our main topic is the interplay of magnetic field and anisotropy. Since the
Zeeman and anisotropy terms in the Hamiltonian do not commute and
are assumed to be comparable in
magnitude, we expect the differential
conductance $g$ to depend significantly on the field. The natural way
to study this is to plot $g$ as a function of quantities characterizing the
magnetic field and possibly of bias voltage. Such \emph{magnetic-field
scans} could, in principle, also be done experimentally.

Since a gate electrode is not required, magnetic-field scans could also be
taken for \emph{monolayers} of molecules sandwiched between conducting
electrodes\cite{ElT07} or with an STM for single molecules or monolayers. For
the theory to be applicable, the tunneling between molecule and electrodes must
be made sufficiently weak to justify the sequential-tunneling approximation,
e.g., by molecular spacer groups or a thin oxide layer.\cite{QNH04} The present
subsection thus applies to single molecules and to monolayers of molecules with
aligned easy axes.

\begin{figure}[t]
\centerline{\includegraphics[width=1.68in]{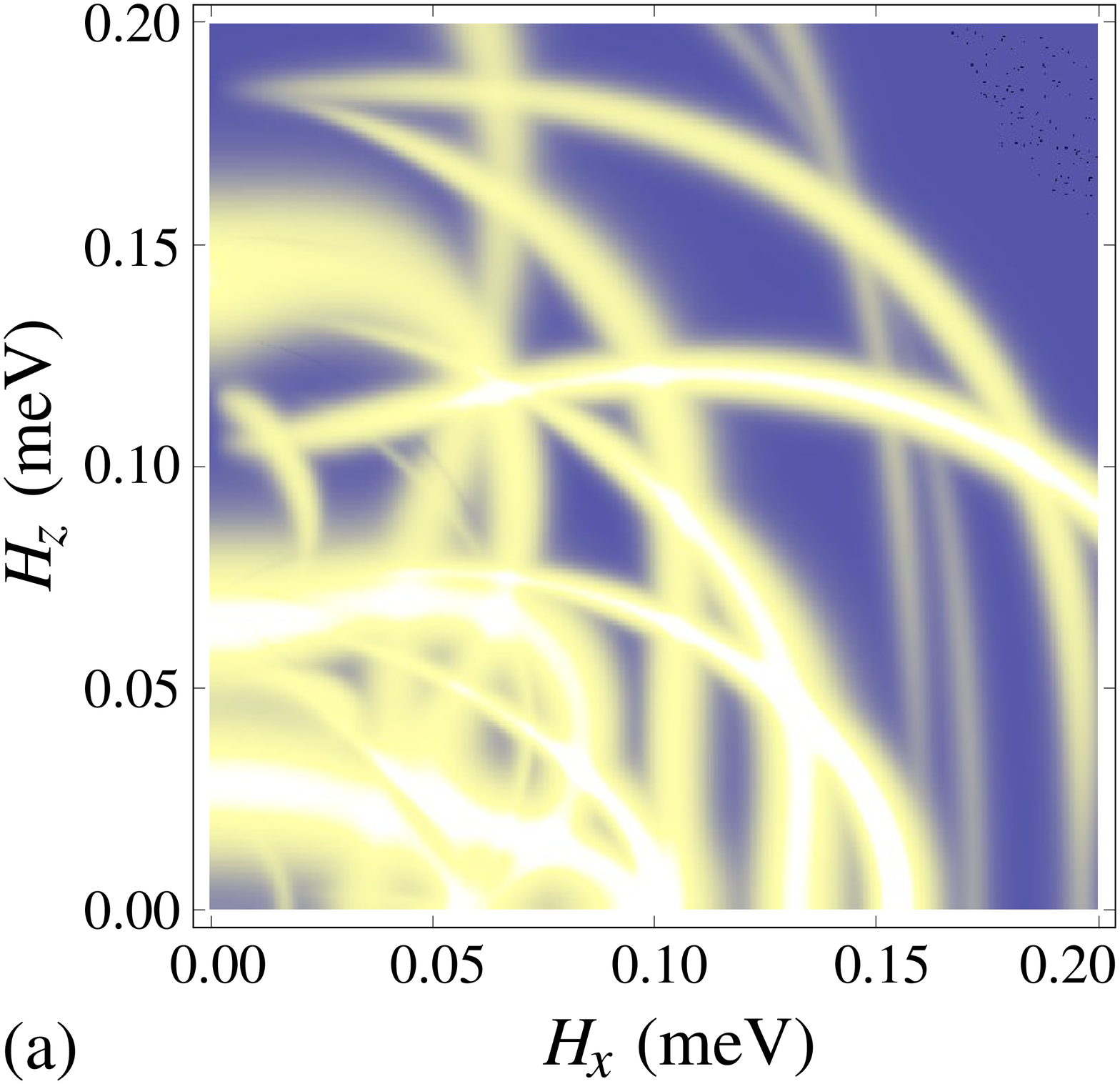}
\includegraphics[width=1.68in]{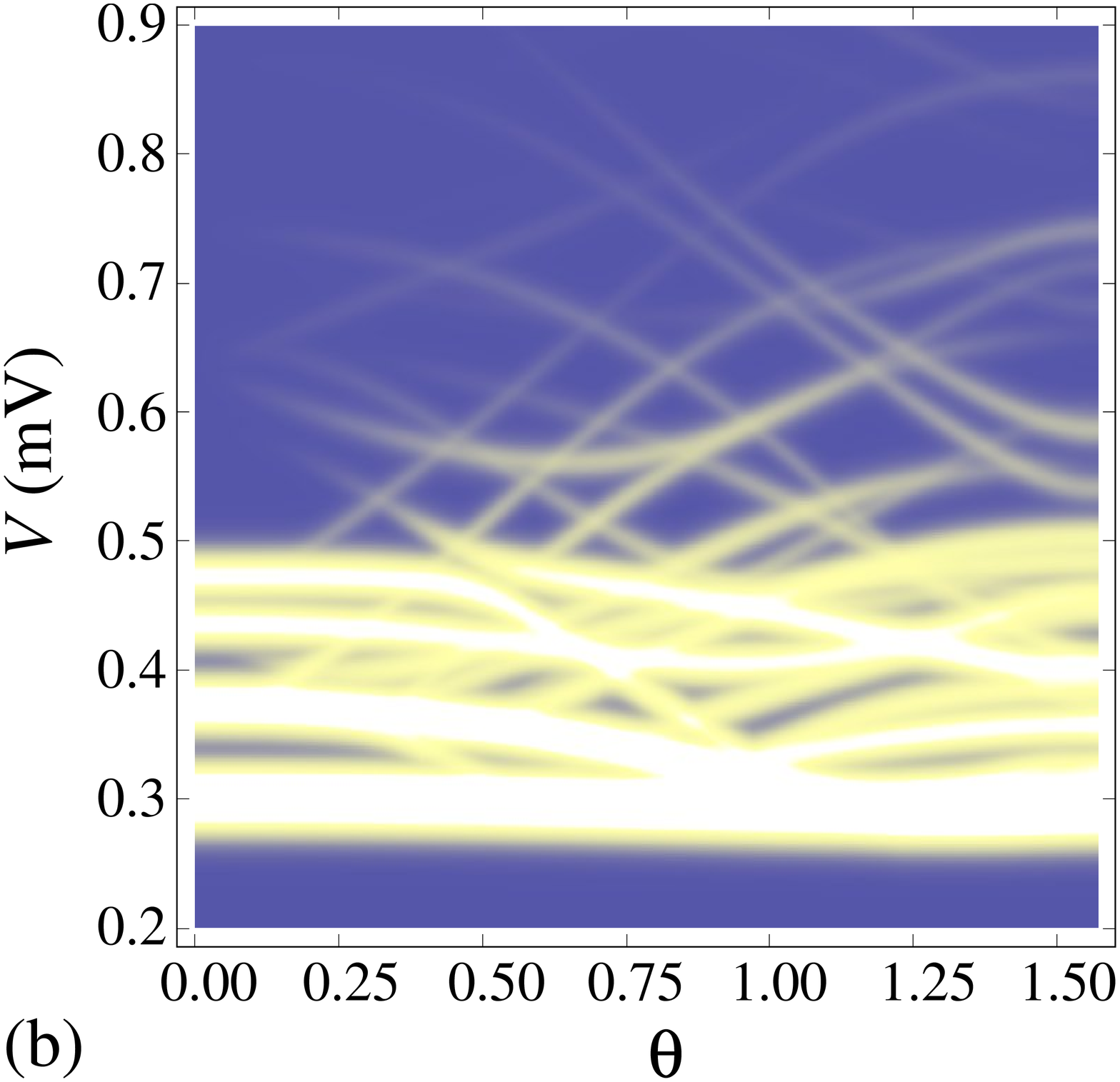}}
\caption{\label{fig.gHH}(Color online) (a) Differential conductance $g$ as a
function of the magnetic field components $H_x$ (perpendicular to easy axis)
and $H_z$ (parallel to easy axis). Bright (dark) colors denote $g>0$ ($g<0$).
The parameters are $S=2$, $\epsilon_0-eV_g=0.2$, $U=10$, $K_2=0.04$,
$\kappa=0$, $J=0.1$, $T=0.002$ (all energies in meV), and $V=0.8\,\mathrm{mV}$.
The rate equations have been solved exactly. (b) $g$ as a function of the angle
$\theta$ between magnetic field and easy axis and bias voltage $V$ for the
$S=2$ model in a magnetic field $|\mathbf{H}|=0.1\,\mathrm{meV}$. The other
parameters are as in Fig.~\protect\ref{fig.g1}. The approximation of
App.~\ref{app.a} has been used.}
\end{figure}


Figure \ref{fig.gHH}(a) shows $g$ as a function of magnetic-field components
$H_x$ and $H_z$ for the case of spin $S=2$ for fixed gate and bias voltages,
$\epsilon_0-eV_g=0.2\,\mathrm{meV}$ and $V=0.8\,\mathrm{mV}$, respectively.
This plot shows $g$ if one fixes $V_g$ and $V$ in Fig.~\ref{fig.g1} and varies
the magnetic field. A peak in $g$ appears whenever an allowed and energetically
possible transition crosses the energy $eV/2$. Figure \ref{fig.gHH}(a) can be
extended to all values of $\mathbf{H}$ using the rotational symmetry of $g$
around the \textit{z} (easy) axis and the reflection symmetry in the
\textit{xy} plane.

Clearly, the symmetry of $g(\mathbf{H})$ allows to \emph{determine} the
orientation of the molecule(s) from a \emph{transport} measurement in a
magnetic field. This is useful since the orientation is typically poorly
controlled and is not easy to determine in break-junction and electromigration
experiments.


The structure in Fig.~\ref{fig.gHH}(a) is richer than that in
Fig.~\ref{fig.g1}: Curves of large $g$ are not straight and they vary in
intensity (height of the current step) and can even vanish. They are not
straight because they correspond to differences of two eigenenergies of
$H_{\mathrm{mol}}$ and the eigenenergies are complicated functions of
$\mathbf{H}$, since the Zeeman term does not commute with the anisotropy term.
Nevertheless it is remarkable that the relatively simple model with spin
$S=2$ generates this complex structure.

The $g$ peaks in Fig.~\ref{fig.gHH}(a) change in intensity with magnetic field
since the transition matrix elements $C^\sigma_{mn}$ change. In the limiting
case of $H_x=0$, i.e., field parallel to the easy axis, the Zeeman and
anisotropy terms do commute. In this case $S_{\mathrm{tot}}^z$ is conserved,
and transitions changing $S_{\mathrm{tot}}^z$ by values other than $\pm 1/2$
are forbidden. Curves belonging to these transitions vanish for $H_x\to 0$.

Figure \ref{fig.gHH}(b) shows $g$ as a function of bias voltage $V$ and the
angle $\theta$ between field and easy axis. Note again that several peaks
vanish for $\theta\to 0$, where the transitions become forbidden. The plot
shows that the strong peak at the CB threshold also depends on the
magnetic-field direction. It is thus possible to switch the molecule between
CB, with very small current due to cotunneling,\cite{ElT07} and a state with
large current, leading to a large and anisotropic magnetoresistance at
low temperatures.

\begin{figure}[t]
\centerline{\includegraphics[width=3.20in]{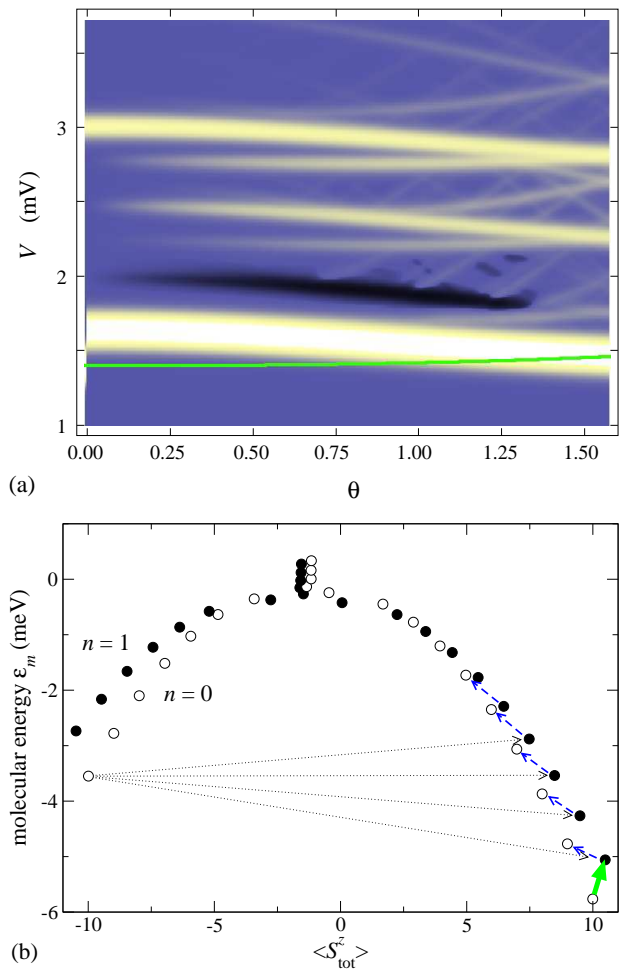}}

\vspace*{1.5ex}
\centerline{\includegraphics[width=3.20in,clip]{timmfig7b.eps}}
%
\caption{\label{fig.gthetaV2}(Color online) (a) Differential conductance
$g$ as a function of the angle
$\theta$ between magnetic field and easy axis and bias voltage $V$ for
$\mathrm{Mn}_{12}$ in a magnetic field $|\mathbf{H}|=0.116\,\mathrm{meV}$
($1\,\mathrm{T}$) at temperature $T=0.00863\,\mathrm{meV}$ ($0.1\,\mathrm{K}$).
The other parameters are as in Figs.~\protect\ref{fig.g2} and
\protect\ref{fig.g3}.
The approximation of App.~\ref{app.a} has been used.
The solid (green) curve denotes the CB threshold, which is invisible in $g$ for
$\theta>0$ due to the presence of a blocking state.
(b) Molecular energy levels vs.\ spin expectation value $\langle
S_{\mathrm{tot}}^z\rangle$ for $\theta=0.3$.
Open (solid) circles correspond to states with $n=0$ ($n=1$) electrons.
The blocking state has $n=0$ and $\langle S^z_{\mathrm{tot}}\rangle\approx -10$.
Transitions discussed in the text are marked by arrows
(most allowed transitions are \emph{not} marked).}
\end{figure}


Figure \ref{fig.gthetaV2}(a) shows $g(\theta,V)$ for $\mathrm{Mn}_{12}$. On
first glance, similar features as in Fig.~\ref{fig.gHH}(b) are seen.
However, an important new effect is at work here: Analysis of the $\theta=0$
case shows that the CB threshold for $\theta=0$ corresponds to the transition
from $m=10$ to $m=21/2$, where now $m$ is the quantum number of
$S^z_{\mathrm{tot}}$. The corresponding bias voltage is $V_{\mathrm{CB}} =
1.41\,\mathrm{mV}$. However, Fig.~\ref{fig.gthetaV2}(a) does not show a strong
peak at this bias for small $\theta>0$. Rather, the threshold \emph{appears}
to be at about $V = 1.64\,\mathrm{mV}$.

The molecular levels shown in Fig.~\ref{fig.gthetaV2}(b) help to understand
this situation: the first possible transition at $T=0$ is the one from $\langle
S^z_{\mathrm{tot}}\rangle \approx 10$ to $\langle S^z_{\mathrm{tot}}\rangle
\approx 21/2$, denoted by a solid curve in Fig.~\ref{fig.gthetaV2}(a) and a
solid arrow in (b). For $\theta=0$ this leads to a large current associated
with repeated transitions between these two states, which are the only allowed
ones until the transition to $\langle S^z_{\mathrm{tot}}\rangle \approx 19/2$
becomes possible at higher bias. However, for $\theta>0$ there are nonzero
matrix elements for transitions that decrease $\langle
S^z_{\mathrm{tot}}\rangle$ by about $3/2$ (dashed arrows). Consequently, the
molecule can reach the states with $\langle S^z_{\mathrm{tot}}\rangle \approx
0$. From there, it can relax to states with \emph{negative} $\langle
S^z_{\mathrm{tot}}\rangle$ through transitions with large rates, in particular
to the state with $\langle S^z_{\mathrm{tot}}\rangle \approx -10$. But this is
essentially a \emph{blocking} state, since the transitions to $\langle
S^z_{\mathrm{tot}}\rangle \approx -21/2$ and $-19/2$ are energetically
impossible and the only transitions that are energetically possible correspond
to very large changes of $\langle S^z_{\mathrm{tot}}\rangle$ (dotted arrows)
and have extremely small matrix elements. This is thus an example of \emph{spin
blockade}. Note that \emph{cotunneling} transitions out of the blocking state
to $\langle S^z_{\mathrm{tot}}\rangle \approx -9$ are possible, since the full
potential difference $eV$ is available for excitations. This leads to a very
small current dominated by cotunneling, which would still be invisible in
Fig.~\ref{fig.gthetaV2}(a).


To summarize, while transitions from positive to negative $\langle
S^z_{\mathrm{tot}}\rangle$ happen only with a small rate, transitions in the
opposite direction occur with a still \emph{much smaller} rate. Consequently,
the molecule is nearly always in the $\langle S^z_{\mathrm{tot}}\rangle \approx
-10$ state and the current is extremely small.\cite{footnote.dyna} Thus there
is no visible differential-conductance peak at the CB threshold.

We conclude with two remarks: (i) The discontinuity of $g$ at $\theta=0$ is
irrelevant in practice, since it is impossible to perfectly align the molecular
easy axis with the field. (ii) Figure \ref{fig.gthetaV2}(a) shows that the CB
threshold coincides with the strong $g$ peak for $\theta=\pi/2$. In this case,
Fig.~\ref{fig.gthetaV2}(b) would be symmetric under spin inversion so that the
states with $\langle S^z_{\mathrm{tot}}\rangle \approx \pm 10$ become
degenerate ground states and there is no blocking state. This case was studied
in Figs.~\ref{fig.g2} and \ref{fig.g2a}.

\subsection{Orientational disorder}
\label{sub.dis}

So far we have considered tunneling through single mo\-le\-cules or through
monolayers of molecules with parallel easy axes.\cite{FCH05} Depending on the
ligands, $\mathrm{Mn}_{12}$ can be nearly spherical. In that case the
distribution of orientations in a monolayer will be closer to random.


To describe tunneling through a monolayer of randomly oriented magnetic
molecules, we have to average the current and differential conductance over all
orientations. Equivalently, we here average over all magnetic-field directions
relative to the molecular easy axis. One might expect that this smears out most
of the fine structure. As we shall see, this is not the case. The averaged
differential conductance $\overline g$ as a function of bias voltage $V$ is
rather complex due to \emph{van Hove singularities} coming from extrema and
crossings in the transition energies as functions of magnetic field
direction.

The angle-averaged differential conductance per molecule is $\overline g =
(1/4\pi) \int d\theta\,d\phi\,(\sin\theta)\, g$, where $\theta$, $\phi$ are the
polar angles of the field. $g$ shows peaks at bias voltages $V$ corresponding
to molecular-transition energies that depend on the angles. Compared to the
conventional van Hove singularities of band theory, $\overline g$ corresponds
to the density of states, $V$ to the energy, and $(\theta, \phi)$ to the wave
vector. If the transition energies depended on both $\theta$ and $\phi$, they
would form a two-dimensional band structure in a ``Brillouin zone'' that is the
surface of a sphere. Extrema in the transition energies would then lead to
typical two-dimensional, step-like van Hove singularities. However, in our
model the transition energies are independent of $\phi$ so that $\overline g =
(1/2) \int d\theta\,(\sin\theta)\, g$. We obtain a one-dimensional band
structure with an additional weight factor $\sin\theta$.

For an extremum at $0<\theta<\pi$, this leads to a one-sided singularity of the
form $1/\sqrt{|V-V_c|}$, typical for one-dimensional systems. If the extremum
is at $\theta=0$ (and $\theta=\pi$) and the transition is \emph{allowed} there,
the factor $\sin\theta$ reduces the singularity to a step. If the transition is
\emph{forbidden} for $\theta=0$, the peak height in $g$ is found to
vanish as $\theta^2$, which leads to an even weaker singularity linear in
$|V-V_c|$. There are other cases of van Hove singularities, which are not
present in normal band structures and will be discussed below.

It is the calculation of \emph{angle-averaged} quantities for which the
approximation scheme of App.~\ref{app.a} becomes crucial to reduce the
computational effort. All averaged quantities are calculated with this method.
To approximate
\be
\overline g = \frac{1}{2} \int_0^\pi d\theta\,(\sin\theta)\,g
  = \int_0^1 du\, g(u,V) ,
\ee
where $u=\cos\theta$, we first calculate $g$ for fixed $u$ at $T=0$. We
obtain a set of bias voltages $V_i(u)$ and delta-function weights $g_i(u)$,
which depend on $u$. We replace the integral by a sum over
typically $N=5000$ terms,
\be
\overline g \approx \frac{1}{N} \sum_{n=1}^N \sum_i
  g_i({n}/{N})\:
  \delta\!\left(V-V_i({n}/{N})\right) .
\ee
To obtain approximate results for $T>0$ we broaden the delta functions (or
current steps) as described in App.~\ref{app.a}.

\begin{figure}[t]
\centerline{\includegraphics[width=3.20in]{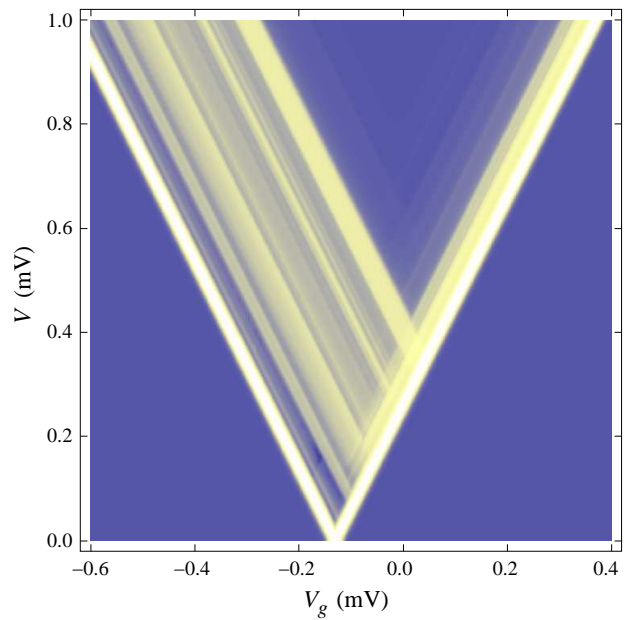}}
\caption{\label{fig.ga1}(Color online) Differential conductance per molecule
averaged over all magnetic-field directions, $\overline g$, as a function of
gate voltage $V_g$ and bias voltage $V$. Bright (dark) colors denote $\overline
g>0$ ($\overline g<0$). The parameters are identical to
Fig.~\protect\ref{fig.g1},
except that the magnetic-field direction is not held fixed but is averaged over,
and the same color scheme is used.}
\end{figure}

For orientation, we first show $\overline g(V_g,V)$ for $S=2$ in
Fig.~\ref{fig.ga1}. The plot should be compared to Fig.~\ref{fig.g1} for the
same parameters except for fixed $\theta=45^\circ$ in Fig.~\ref{fig.g1}. Figure
\ref{fig.ga1} shows that a lot of the fine structure survives the angular
averaging. Plots of this type could not be obtained experimentally, due to the
lack of a gate electrode for a monolayer. What \emph{can} be measured is the
differential conductance $\overline g$ as a function of bias voltage for given
onsite potential $\epsilon_0$. Large signals are expected if the particular
molecule allows to reach the CB threshold with accessible bias voltages $V$.

\begin{figure}[t]
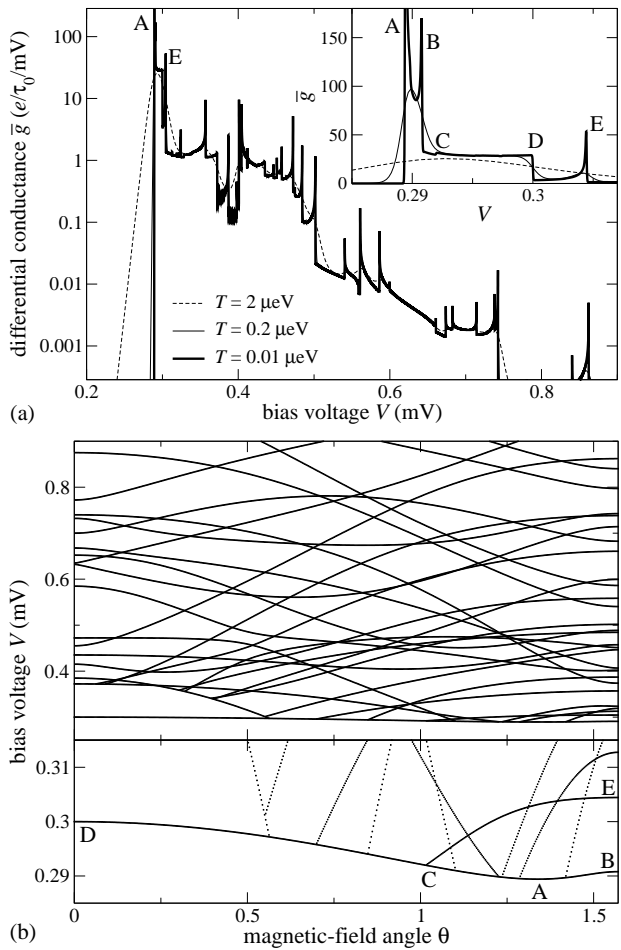

\centerline{\includegraphics[width=3.20in,clip]{timmfig9a.eps}}

\vspace*{1ex}
\centerline{\includegraphics[width=3.20in,clip]{timmfig9b.eps}}
%
\caption{\label{fig.ga2}(a) Differential conductance per molecule
averaged over all magnetic-field directions, $\overline g$, as a function
of bias voltage $V$ at three temperatures.
$\overline g$ is plotted on a logarithmic scale.
The parameters are as in Fig.~\protect\ref{fig.gHH}(b). The inset shows
$\overline g$ in
the vicinity of the CB threshold on a linear scale.
Several van Hove singularities are marked.
(b) Upper part: bias voltage of $g$ peaks as a function of
magnetic-field angle $\theta$, for the same parameters. Lower part:
close-up of the vicinity of the CB threshold. The points corresponding to the
van Hove singularities in (a) are marked by the same letters.}
\end{figure}

Figure~\ref{fig.ga2}(a) shows $\overline g(V)$ for a monolayer of spin
$S=2$ molecules in a magnetic field of magnitude
$|\mathbf{H}|=0.1\,\mathrm{meV}$. This plot corresponds to
Fig.~\ref{fig.gHH}(b) averaged over $\theta$ with the weight factor
$\sin\theta$. We again see that the averaged differential conductance retains
a lot of structure, in particular at very low temperatures. In the following,
we analyze this structure in terms of van Hove singularities.

To help with this analysis, Fig.~\ref{fig.ga2}(b) shows the bias voltages of
$g$ peaks (current steps) as a function of magnetic-field angle $\theta$. 
These are the ``bands'' that lead to van Hove singularities.
This is simply a map of $g$ peaks occuring in Fig.~\ref{fig.gHH}(b). The first,
strong peak in Fig.~\ref{fig.ga2}(a) corresponds to the CB threshold. Figures
\ref{fig.gHH}(b) and \ref{fig.ga2}(b) show that the CB threshold is
angle-dependent. Singularity A stems from the minimum of the transition energy
and B from the maximum at $\theta=\pi/2$. These are both of the typical
one-dimensional form $1/\sqrt{|V-V_c|}$. The maximum at $\theta=0$ leads to
singularity D, which is only  a step, due to the weight factor
$\sin\theta$. As noted above, the presence of a step shows that the transition
is allowed for $\theta = 0$. The $1/\sqrt{|V-V_c|}$ singularity E
stems from the maximum of a transition not at the CB threshold.

In addition, there is weaker step C, which cannot be attributed to an extremum
at $\theta=0$. Figure \ref{fig.ga2}(b) shows that a transition can suddenly
vanish when it intersects another one. This is a typical property of inelastic
sequential tunneling through molecules, which is due to the initial state of
one of the transitions being populated (at $T=0$) only on one side of the
crossing. One can show that the \emph{sum} of weights (current step heights) of
the transitions is analytic through the crossing, as a function of $H_x$,
$H_z$, $\theta$, or $V_g$.\cite{footnote.continuity} This means that a
three-way crossing can be viewed as the superposition of a band for which the
energy and the weight are analytic functions of $\theta$ and another band for
which the weight is analytic but the energy shows a kink (change of slope).
This kink leads to a type of van Hove singularity not present in band
structures for weakly interacting electrons in solids. It shows up as a
kink in the current and thus as a step in $\overline g$.
Singularity C in Fig.~\ref{fig.ga2}(a) is of this type, coming from the
three-way crossing marked in Fig.~\ref{fig.ga2}(b). If both
transitions have nonzero weight on both sides, part of the weight is also
typically transferred from one band to the other, leading to the same type of
singularity.


\begin{figure}[t]
\centerline{\includegraphics[width=3.20in,clip]{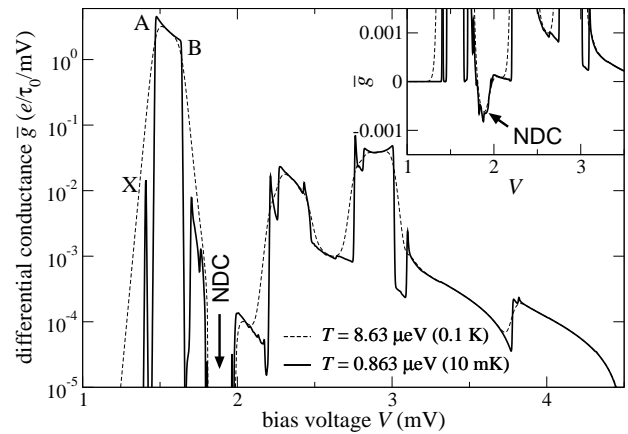}}
\caption{\label{fig.ga3}Differential conductance per molecule
averaged over all magnetic-field directions, $\overline g$, as a function
of bias voltage $V$ for a monolayer of $\mathrm{Mn}_{12}$ molecules with
randomly oriented easy axes at
two temperatures. $\overline g$ is plotted on a logarithmic scale.
The inset shows $\overline g$ on a linear scale to exhibit the NDC region.
Several van Hove singularities are marked, see text.}
\end{figure}

Finally, we turn to a monolayer of $\mathrm{Mn}_{12}$ molecules ($S=10$) with
random easy axes. We consider the case corresponding to averaging $g$ in
Fig.~\ref{fig.gthetaV2}(a) over angles. The result is the differential
conductance plotted in Fig.~\ref{fig.ga3}. The most prominent features are the
step-like singularities A and B. B stems from the maximum of the transition
energy of the strong low-bias peak in Fig.~\ref{fig.gthetaV2}(a) at about $V =
1.64\,\mathrm{mV}$. It is a step since the maximum occurs at $\theta=0$.
However, singularity A is also a \emph{step}, although it clearly results from
the minimum of that transition energy at $\theta=\pi/2$ so that we expect a
\emph{pole}, $1/\sqrt{|V-V_c|}$. This can be understood by extending
Fig.~\ref{fig.gthetaV2}(a) to angles $\theta>\pi/2$ using reflection symmetry.
The minimum is in fact a \emph{crossing} between the transitions from $\langle
S^z_{\mathrm{tot}}\rangle = \pm 10$ to $\pm 21/2$, i.e., the transitions out of
the respective ground state and blocking state. Therefore singularity A is not
of the form of a band extremum but of a band crossing, i.e., a step. The
anomalous exponent of singularity A is thus closely related to the spin
blockade discussed earlier.

The sharp peak X in Fig.~\ref{fig.ga3} is an artifact coming from molecules
with $\theta=0$, for which the CB threshold formally occurs at $V_{\mathrm{CB}}
= 1.41\,\mathrm{mV}$. Finally, the extended region of NDC in
Fig.~\ref{fig.gthetaV2}(a) leads to NDC even in the angle-averaged differential
conductance, as seen in the inset of Fig.~\ref{fig.ga3}.


\section{Summary and conclusions}
\label{sec.sum}

In summary, we have studied inelastic electron tunneling through molecules with
a local magnetic moment and large uniaxial anisotropy in a strong magnetic
field. Since the orientation of the molecules is often not well controlled in
tunneling experiments, we consider arbitrary angles between easy axis and
field. Then the anisotropy and Zeeman terms in the Hamiltonian do not
commute so that \emph{no} component of the molecular spin is conserved. This
lifts all spin selection rules for electron tunneling, leading to large numbers
of allowed molecular transitions and consequently to many peaks in the
differential conductance $g=dI/dV$. The resulting complex fine structure of
Coulomb-diamond plots is already apparent for the relatively simple case of a
local spin $S=2$.

As a concrete example, $\mathrm{Mn}_{12}$ molecules are studied. The large spin
$S=10$ leads to even more complex fine structure. However, one can still
attribute  differential-conductance peaks to individual molecular transitions.
It should be possible to analyze experimental results in terms of specific
molecular transitions \emph{if} one has a model for guidance. Even without
detailed attribution of observed peaks, measurement of $g$ in magnetic fields
of various directions should allow to determine the orientation of the molecule
relative to the leads, which is not directly accessable in break-junction or
electromigration experiments.

We have considered three cases: Single molecules, molecular monolayers with
aligned easy axes, and molecular monolayers with random easy axes. For
monolayers one does not have the advantage of a gate electrode. However, one
can extract similarly detailed information by varying the magnetic field. For
randomly oriented molecules we have found that the averaging of transition
energies over orientations leads to van Hove singularities in $g$. Besides the
normal singularities from extrema of ``bands,'' a novel type arises from
crossings of transition energies. Analysis of the
singularities can give rather detailed information on allowed vs.\ forbidden
transitions in the limit of the easy axis aligned with the magnetic field and
on the presence of blocking states (spin blockade).

Detailed calculations of $g$ for $\mathrm{Mn}_{12}$ with its many molecular
states and in particular for monolayers with random orientation are made
feasible by an approximation scheme for solving the rate equations. As further
results, we predict large and highly anisotropic magnetoresistance at low
temperatures, if the bias voltage is tuned close to the CB threshold, and
negative differential conductance, which survives even for randomly oriented
monolayers of $\mathrm{Mn}_{12}$.



\acknowledgments

The author would like to thank F. Elste for useful discussions and the Kavli
Institute for Theoretical Physics, Santa Barbara, for hospitality while part of
this work was performed. This research was supported in part by the National
Science Foundation under Grant No.\ PHY99-07949.

\appendix

\section{Approximate solution of the rate equations}
\label{app.a}

At $T=0$ we employ the following algorithm: For bias $V=0$, the molecular
ground states have probability $1/d$, where $d$ is the ground-state degeneracy,
and the current vanishes. For increasing $V$, the current remains zero in the
sequential-tunneling approximation until $eV/2$ reaches the  energy
$\Delta\epsilon$ of the first allowed transition starting from a ground state.
Above this value, all states obtain nonzero probabilities that can be reached
from a ground state directly or indirectly by allowed and energetically
possible transitions. The network of these states is constructed by testing all
possible transitions. Then the matrix $A$ above the threshold
$eV/2=\Delta\epsilon$ is obtained from Eq.~(\ref{C1.R4}), taking into account
that the Fermi factors are $f=0$ or $1$. $A$ is then diagonalized. This is
faster and more robust than in the general case because (a) the dimension of
$A$ is smaller since it only contains active states and (b) $A$ does not
contain exponentially small components---all components are either exactly zero
or given by matrix elements. The resulting current is
calculated from Eq.~(\ref{C1.I5}). The value of $\Delta\epsilon$ and the change
(step height) in the current are recorded. Then we go over to the next bias
for which $eV/2$ equals a transition energy, extend the network of states
by all states that can now be reached, obtain $A$, diagonalize it, and
calculate the new current. These steps are repeated.

Note that we only diagonalize $H_{\mathrm{mol}}$ \emph{once} to obtain the
transition energies and matrix elements. Furthermore, we only diagonalize $A$
\emph{once for each transition energy}, since the current at $T=0$ is constant
between steps. The result is a list of bias voltages $V_i$ and associated
heights $g_i$ of current steps. The $g_i$ are also the weights of
delta-function peaks in $g=dI/dV$.



To obtain approximate results at finite, but low, temperatures, we broaden
the current steps so that
their width is of the order of $k_BT$. We write the current as
\be
I \approx e\sum_i g_i\, \left[ f\!\left(\frac{eV_i -eV}{2k_BT}\right)
  - f\!\left(\frac{eV_i + eV}{2k_BT}\right) \right] ,
\label{2.tI2}
\ee
where $i$ enumerates the current steps.
This particular broadening function is chosen
since Eq.~(\ref{2.tI2}) is \emph{exact} (assuming sequential tunneling) for the
simplest possible model, i.e., a single orbital for a spin-less fermion.
The form of the broadened delta functions in $g$
follows trivially. The approximation is good if the separation in
$eV/2$ between current steps is large compared to $k_BT$.




\begin{thebibliography}{99}
\bibitem{footnote.highmulti}The higher-energy multiplet of states with $n=1$,
which is split from the lower multiplet by an energy of order $JS$, also
becomes occupied due to tiny matrix elements between $n=0$ states with large
negative $langle S_{\mathrm{tot}}^x\rangle$ and $n=1$ states with large
positive  $langle S_{\mathrm{tot}}^x\rangle$. However, the probabilities are
negligible for the present discussion.
\bibitem{footnote.Jstates}There is another multiplet of $n=1$ states at higher
energies, split from the low-energy multiplet by a large energy of the order of
$JS$. These states are not populated for the bias voltages considered here.
\bibitem{footnote.continuity}This follows because the current at $T=0$ is
analytic, as long as no transition is crossed, since the Fermi functions are
then constant and the matrix elements are analytic functions of magnetic field
and gate voltage.
\end{thebibliography}
\end{document}